\documentclass[a4paper,11pt]{article}

\usepackage{jheppub}
\usepackage[usenames,dvipsnames,table]{xcolor}
\usepackage{graphicx,amsmath,amssymb,multirow,array,bm,mathrsfs,bbold,bbm}
\usepackage{appendix}
\usepackage{footmisc}
\usepackage{epsf,amsfonts,slashed,soul}
\usepackage{lipsum}
\usepackage{subfig}
\usepackage{pifont}
\usepackage{bbold}
\usepackage{dsfont}
\usepackage{tikz}
\usepackage{verbatim}
\usepackage{placeins}
\usepackage{physics}
\usepackage{graphicx}
\usepackage{comment}
\usepackage{xcolor}

\def\ie{i.e.~}

\newcommand{\cS}{{\mathcal{S}}}

\newcommand{\be}{\begin{equation}}
\newcommand{\ee}{\end{equation}}
\newcommand{\bea}{\begin{eqnarray}}
\newcommand{\eea}{\end{eqnarray}}

\begin{document}

\title{Islands in FRW Cosmologies
}

\author{Ricardo Esp\'indola,}
\author {Bahman Najian,}
\author{and Dora Nikolakopoulou}
\affiliation{Institute for Theoretical Physics, University of Amsterdam, Science Park 904, Postbus 94485, 1090 GL Amsterdam, The Netherlands}

\emailAdd{r.espindolaromero@uva.nl}
\emailAdd{b.najian@uva.nl}
\emailAdd{t.nikolakopoulou@uva.nl}

\abstract{We search for candidate island regions in FRW cosmologies supported by radiation, cosmological constant $\Lambda$, and non-zero spatial curvature. The radiation is assumed to be in a thermal state. We apply the necessary conditions introduced in \cite{Hartman:2020khs}.  Both for the open and closed universes with $\Lambda<0$, the Friedmann equation admits recollapsing solutions with a time-symmetric slice. In the case of closed universes, there is always an island that is the whole Cauchy slice. However, for $\Lambda<0$, we also find another finite-sized candidate island region, in the middle of the spacetime, at the turnaround time. In the case of the open universes, we only find a candidate island region for $\Lambda<0$, that appears at the turnaround time. It starts from a finite value of the radial coordinate and extends to infinity. Looking at both open and closed universes, we conclude that the key ingredient that allows the existence of islands is the negative cosmological constant. We provide analytic results along the time-symmetric slices and support our results with numerics in the whole spacetime.}
\maketitle
\section{Introduction}
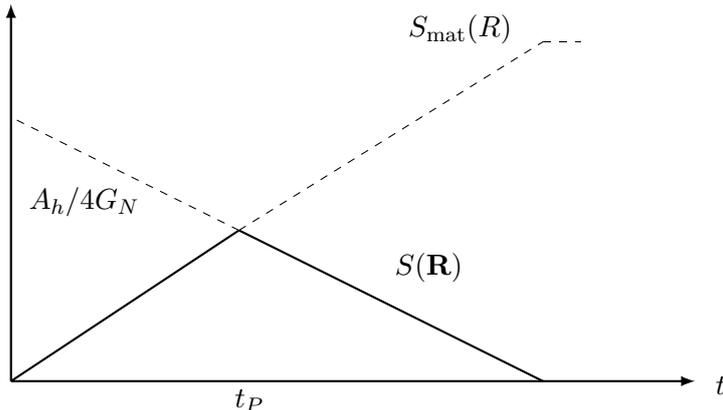
\begin{figure}[t]
    \centering
    \begin{tikzpicture}
    \tikzset{>=latex}
    \draw[ thick,->] (-4,0) -- (5,0);
    \draw[thick,-] (-4,0) -- (-1,2);
    \draw[thick, -] (-1,2) -- (3,0);
\draw[thick,->] (-4,0) -- (-4,5);
\draw[dashed] (-1,2) -- (-4,3.5);
\draw[dashed] (-1,2) -- (3,4.5);
\draw[dashed] (3,4.5) -- (3.5,4.5);
\put(150,-5){$t$}
\put(30,40){$S(\mathbf{R})$}
\put(35,130){$S_{\rm mat}(R)$}
\put(-107,65){$A_h/4G_N$}
\put(-30,-10){$t_P$}
\end{tikzpicture}
{\\ \caption{Page curve of the fine-grained entropy of an evaporating black hole.
}
\label{Fig:page}}
\end{figure}

One of the long-standing puzzles in modern physics is the black hole information paradox.
Its essence can be captured by examining the entropy of the sub-systems of an evaporating black hole. In Hawking's seminal calculation \cite{Hawking:1975vcx,Hawking:1976ra}, the fine-grained entropy of the radiation seemingly exceeds the Bekenstein-Hawking entropy of the black hole. This signifies information loss. In a unitary process, a pure state evolves into a pure state. Page showed \cite{Page:1993wv} that initially the fine-grained entropy grows, following the Hawking curve, but approximately halfway through the evaporation process, it starts decreasing and eventually vanishes, which is consistent with unitary evolution. 
In a series of breakthrough papers \cite{Almheiri:2019psf,Penington:2019npb,Almheiri:2019hni, Almheiri:2019qdq, Penington:2019kki,Almheiri:2020cfm}, it was shown that the Page curve can be recovered within semiclassical gravity. The key to this advance was realizing that the fine-grained entropy of the radiation receives contributions from a disconnected region that lies in the gravitating system, namely, the island. The exact entropy of the radiation is given by the island formula 
\be\label{eq:island}
S(\mathbf{R}) = \underset{I}{\text{min}}\Big\{ \underset{I}{\text{ext}} \Big[ \frac{{A}(\partial I)}{4G_N} +  S_{\text{mat}}(R \cup I)\Big] \Big\}\,,
\ee
where $A(\partial I)$ is the area of the boundary of the island $I$, and $S_{\rm mat}(R \cup I)$ is the renormalized entropy of the quantum fields on the union of the regions $R$ and $I$. The formula instructs us to extremize and minimize over all possible islands. Before the evaporation begins, there are not any non-trivial islands and no Hawking pairs have been emitted. So, initially, the exact entropy of the radiation is 
\be
\label{eq:vNrad}
S(\textbf{R})\approx S_{\text{mat}}(R)\,.
\ee
When the process starts, more and more Hawking partners escape from the black hole and \eqref{eq:vNrad} steadily grows. As the evaporation proceeds, a non-trivial island appears in the interior of the black hole. Its boundary is very close to the black hole horizon. Since it extends almost through the whole black hole interior, it contains most of the partners of the Hawking radiation that have escaped from the black hole. The partners contained in the island region purify the ones of the black hole, so the term $S_{\mathrm{mat}}(R\cup I)$ in \eqref{eq:island} vanishes and the exact entropy of the radiation becomes
\be
\label{area}
S(\textbf{R})\approx\frac{A(\partial I)}{4G_N}\,.
\ee
As the black hole horizon shrinks, \eqref{area} decreases and finally vanishes. Thus, the Page curve is followed and the black hole evaporation process is unitary (see Figure \ref{Fig:page}). Remarkably, the island formula (\ref{eq:island}) has been derived using the Euclidean path integral by applying the gravitational replica trick \cite{Lewkowycz:2013nqa} (See \cite{Almheiri:2020cfm} for a review). It was shown that after the Page time replica wormholes become dominant  \cite{Almheiri:2019qdq, Penington:2019kki}.
The island formula \eqref{eq:island} is a generalization of Ryu-Takayanagi formula \cite{Ryu:2006bv, Hubeny:2007xt,Faulkner:2013ana,Engelhardt:2014gca,Dong:2016hjy,Dong:2016fnf} which has been extensively studied in the literature \cite{Almheiri:2019yqk, Almheiri:2019psy,Chen:2019uhq, Rozali:2019day, Chen:2020uac,Balasubramanian:2020coy,Balasubramanian:2020xqf,Bousso:2020kmy,Goto:2020wnk,Bousso:2021sji,Chen:2020tes}.

It is worth emphasizing once more that in \cite{Almheiri:2019qdq, Penington:2019kki} the Page curve was recovered in the context of semiclassical gravity. This leads one to ask whether islands exist in cosmological spacetimes, where we do not have AdS/CFT duality \cite{Maldacena:1997re} to assist us. Moreover, since our universe has positive cosmological constant, it is natural to wonder whether or not we might be living in an island.

The first cosmological islands were found in \cite{Hartman:2020khs}. The authors considered a setup where a radiation-dominated, flat Friedmann-Robertson-Walker (FRW) spacetime is entangled with a non-gravitating auxiliary system. They examined the cases with zero, positive and negative cosmological constant and concluded that islands appear only in the last case. The way they achieved this was by introducing three conditions that aid the search for islands. The beauty of these conditions is that they are independent of the radiation region $R$. Once a non-trivial island region is found, it is natural to wonder where the information of the degrees of freedom in $I$ is encoded. One possible way to answer this question is by introducing an auxiliary system that purifies the state of the system in $I$, for example preparing the whole state in a thermofield double like state. This process is however non-unique.

In this paper, we build upon the work of \cite{Hartman:2020khs}. We extend their analysis to FRW universes with non-zero spatial curvature. We consider a state that is approximately thermal with inverse temperature $\beta$. We use the island conditions proposed in \cite{Hartman:2020khs} along with an extra set of conditions that ensure our candidate islands are in the semiclassical regime. In regions of spacetime where all of the above are satisfied we conclude that islands can exist. Our main results are summarized below.

\textbf{Summary of results}: As was mentioned before, we study FRW cosmologies that are supported by radiation in a thermal state, cosmological constant and non-zero spatial curvature. In closed universes, with any type of cosmological constant, there is always an island that is the whole Cauchy slice. Additionally, when $\Lambda<0$, for any spatial curvature, we find that there is a different type of candidate island region. These universes are recollapsing and have a time-symmetric slice. For $k>0$ and $\Lambda<0$, we find that there is a candidate island region in the middle of the Penrose diagram around the time-symmetric slice. This region is shown in Figure \ref{Fig:Λ<0,k>0}. For $k<0$ and $\Lambda<0$, we have a similar situation, i.e. a candidate island  region around the time-symmetric slice, only this time it starts at a finite value of the radial coordinate and extends to infinity, as shown in Figure \ref{Fig:Λ<0,k<0}. This island is similar to the one found in \cite{Hartman:2020khs} for $k=0$ and $\Lambda<0$. We conclude that the main element that allows for the existence of candidate island  regions is a negative cosmological constant. 

For the purpose of this study, we use both analytical and numerical methods. In order to have analytic control of the solution to the Friedmann equation, in all the universes that it is possible, we focus on the time-symmetric slice, where $a'(\eta)\lvert_{\eta=\eta_0}=0$. We study the solution, $a(\eta)\lvert_{\eta=\eta_0}=a_0$, at the low and high temperature limits and find that for $\Lambda<0$ islands appear only in the latter case, as the former is always in the non-semiclassical regime. In the high temperature limit, we see that $a_0$ does not have contributions from the curvature, i.e., to leading order it coincides with the solution of the flat radiation-dominated FRW universe evaluated at its time-symmetric slice. Hence, we conclude that at the turnaround time, in the high temperature limit, the spatial curvature is negligible. In order to support and complement our analytic calculations, we also ``scan" the whole spacetime for candidate island regions numerically and provide multiple figures that show where the existence of islands is possible. 

\textbf{Outline:} This paper is organized as follows. We begin in
Section \ref{framework} by introducing the setup and general framework.
In Section \ref{flatcase} we review the analysis of the radiation-dominated, flat FRW universes done in \cite{Hartman:2020khs}. In Sections \ref{positivecurvature} and \ref{negativecurvature}, we explore the possibility of islands in radiation-dominated FRW cosmologies with positive and negative curvature respectively. Finally, in Section \ref{Sec:Discussion} we briefly summarize our results and discuss future directions.

{\bf Note added}: While finishing this work, the paper \cite{Bousso:2022gth} appeared on the arXiv which has some overlap with our results.

\section{General framework} \label{framework}
We are interested in finding candidate island regions in four dimensional FRW cosmologies. The metric in conformal coordinates is given by
\be \label{metric}
ds^2 = a^2(\eta) \left( -d\eta^2 + d \chi^2 + S^2_k (\chi) d\Omega_2^2 \right)~, ~~
S_k (\chi) := 
\begin{cases} R_0 \sinh (\chi/R_0)~, & k=-1~ \\ \chi ~, & k=0~, \\
R_0 \sin(\chi / R_0)~, & k=1~ \end{cases}
\ee
for open, flat, and closed universes, respectively. Here, $R_0$ is a fixed length scale and $a(\eta)$ is the scale factor. We assume that the state of the system is approximately thermal
\be
\rho \approx \frac{1}{Z} e^{- \beta H}~, 
\ee
where $\beta =  \beta_0 a(\eta) $ is the inverse temperature at conformal time $\eta$, and $\beta_0$ refers to the temperature in Minkowski spacetime. Given a subregion $I$ on a particular Cauchy slice at time $\eta=\eta_0$, the matter entropy of the bulk fields enclosed in it is given by the thermal entropy
\be\label{eq:smat}
S_{\rm mat}(I) =  s_{\rm th} \widetilde{V}(I)~,
\ee
where $s_{\mathrm{th}}$ is the thermal entropy density, and $\widetilde{V}(I)$ is the comoving volume enclosed by $I$. For the metric (\ref{metric}), the comoving volume is
\be
\widetilde{V}(I) = 4\pi \int\limits_{0}^{\chi_I}d\chi~ S_{k}^2(\chi)~.
\ee
We define the region $G$ as the complement of region $I$ in the gravitating system such that they share the same boundary, $\partial I = \partial G$. The entropy of matter fields enclosed by $G$ is also extensive and is simply
\be
S_{\rm mat}(G) = s_{\mathrm{th}} \left(\widetilde{V}_{\rm total} - \widetilde{V}(I) \right)~,
\ee
where $\widetilde{V}_{\rm total}$ is the total volume enclosing both regions $G$ and $I$. We consider FRW cosmologies supported by radiation, spatial curvature, and cosmological constant. The radiation and entropy densities have the form
\be
\rho_{\mathrm{rad}} = \frac{c_{\mathrm{th}} T_0^4}{a(\eta)^4} \ \ \ ~ {\rm and}~\ \ \  s_{\mathrm{th}} = \frac{3}{4} c_{\mathrm{th}} T_0^3~,
\ee
respectively. Here, $c_{\rm th}$ is proportional to the number of degrees of freedom and $T_0$ is the temperature in Minkowski spacetime. The scale factor $a(\eta)$ is the solution to the Friedmann equation\footnote{For the case of a negative cosmological constant, we will define a positive $\Lambda_0 >0$ such that $\Lambda = - \Lambda_0$.}
\be\label{eq:Friedmann}
\frac{1}{a(\eta)^2}\left(\frac{a'(\eta)}{a(\eta)}\right)^2 = \frac{8 \pi G_N}{3}\frac{ c_{\mathrm{th}} T_0^4}{a(\eta)^4} - \frac{k }{a(\eta)^2 R_0^2} + \frac{\Lambda }{3}\,.
\ee
We will apply the necessary island conditions for the existence of islands \cite{Hartman:2020khs}. The explicit form of these three conditions in spacetimes arising from (\ref{metric}) are as follows

\begin{itemize}
\item The Bekenstein bound is violated:
\be\label{eq:BB}
\widehat{S}_{\rm mat}(I) \gtrsim \frac{A(\partial I)}{4 G_N}~,
\ee
where $\widehat{S}_{\rm mat}$ is the finite part of the matter entropy and the wiggly inequality means that the subleading terms compared to right hand side are ignored. The derivation of this condition needs a careful treatment of UV divergences of the matter entropy \cite{Hartman:2020khs}. In our case, the thermal entropy (\ref{eq:smat}) represents the extensive part of $\widehat{S}_{\rm mat}(I)$ and is finite.
\item $I$ is a quantum normal region:
\be\label{eq:Inormal}
\left(\pm \partial_\eta + \partial_{\chi} \right) S_{\rm gen}(I) \geq 0~.
\ee
\item $G$ is a quantum normal region:
\be\label{eq:Gnormal}
\left(\pm \partial_\eta - \partial_{\chi} \right) S_{\rm gen}(G) \geq 0~.
\ee
\end{itemize}
Notice that in both conditions, we are using the null directions to deform the surfaces $\partial I$ and $\partial G$ with respect to the region $I$. For a closed universe, for example, the entire Cauchy slice always satisfies these three conditions.

Once an overlapping region is found in a FRW cosmology, we still need to be sure that they live in the semiclassical regime. In order to do so, we use the following semiclassical conditions:
\begin{itemize}
\item The proper time to the singularity requires
\be\label{eq:time}
\Delta \tau = \int\limits_0^\eta d \eta \ a(\eta) \gg l_P~.
\ee

\item The thermal length scale should satisfy
\be\label{eq:beta}
\beta \gg l_P~, ~~~ \mathrm{\ie} ~~~ \frac{a(\eta)}{T_0} \gg l_P~. 
\ee
\item The energy density should satisfy
\be\label{eq:energy}
\rho_{\mathrm{rad}}=\frac{c_{\mathrm{th}}T_0^4}{a(\eta)^4} \ll M_P^4~.
\ee
\item The size of the $S^2$ obeys
\be\label{eq:sphere}
a(\eta) S_{k}(\chi) \gg l_P~. 
\ee
\item The curvature radius requires
\be\label{eq:curvature}
a(\eta) R_0\gg l_P~.
\ee
\end{itemize}
Let us emphasize that the island conditions together with the semiclassical conditions give a strong indication that an island exists in a given spacetime without making any reference to the radiation region $R$.

Consider the case when we find a candidate island region $I$, \ie, a region in spacetime that fulfills the above criteria. A natural question to ask is in which auxiliary system, $R$, is this region encoded? 
Following \cite{Hartman:2020khs}, one possible way to answer this question is by purifying the original thermal state with a second copy of Minkowski space and preparing the whole system in the thermofield double state (TFD) using the Euclidean path integral
\be 
 |\beta_0  \rangle = \frac{1}{\sqrt{Z}} \sum_{n} e^{-\beta_0 E_{n}}|n\rangle_{1}^{*}|n\rangle_{2}~.
\ee
This is of course a non-unique procedure. In this paper, we are agnostic about the radiation region $R$ and focus more on the question of whether islands can exist in FRW cosmologies. 

\section{Flat slicing} \label{flatcase}
In this section, we review the results of \cite{Hartman:2020khs} for FRW cosmologies with flat slicing. 
\subsection{Zero cosmological constant} We first consider the case where the vacuum energy density is zero. The solution of the Friedmann equation (\ref{eq:Friedmann}) has the simple form. It is given by
\be
a(\eta) =\sqrt{\frac{8 \pi G_N c_{\mathrm{th}} T_0^4}{3}} \eta~.
\ee

\subsection*{Island conditions}
For a spherical region $I$ located at $\chi_I$ at time $\eta_I$, the Bekenstein bound is violated for
\be
\frac{\chi_I}{\eta_I} \gtrsim \frac{3 \pi}{2}T_0 \eta_I~.
\ee
In order for region $I$ to be quantum normal, the ingoing condition is true for values
\be
  \frac{\chi_I}{\eta_I} \geq
    \begin{cases}
      \frac{\pi T_0 \eta_I}{\pi T_0 \eta_I-1}~, & \text{}\ \pi T_0 \eta>1 \\
      0~, & \text{}\ \pi T_0 \eta<1~,
    \end{cases}
\ee
while the outgoing condition is always satisfied. The third condition implies that the outward part is true for
\be
\frac{\chi_I}{\eta_I} \geq \frac{\pi T_0 \eta_I}{\pi T_0 \eta_I+1}~,
\ee
whereas the ingoing, for $\pi T_0 \eta<1$, is satisfied when
\be
\frac{\chi_I}{\eta_I} \geq  \frac{\pi T_0 \eta_I}{1-\pi T_0 \eta_I}~.
\ee
There is an overlapping region where the three conditions are fulfilled for values of conformal time such that 
\be
\label{regnonsem}
T_0 \eta_I < \frac{1}{\pi}~.
\ee 

\subsection*{Semiclassical regime}
Using the proper time condition (\ref{eq:time}), we find that we are in the semiclassical regime when
\be
T_0 \eta_I \gg \frac{1}{c_{\mathrm{th}}^{1/4}}~,
\ee
which is in conflict with \eqref{regnonsem}.

\subsection*{Conclusion}
\begin{figure}[t]
 \centering
     \includegraphics[width=.5\linewidth]{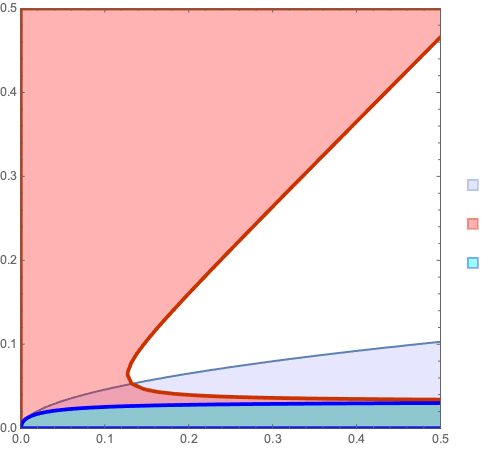}
      \begin{picture}(0,0)
\put(0,113){Bekenstein violating}
\put(0,97){$I$ quantum normal}
\put(0,78){$G$ quantum normal}
\put(-230,100){$\eta$}
\put(-122,-6){$\chi$}
\end{picture}
     \caption{
     Regions where the three island conditions are satisfied. We chose the numeric values $k=0$, $\Lambda=0$, $c_{th}=1$, $T_0=10$, and $G_N=0.01$. There is an overlapping region close to $\eta \approx 0$. However, this region is outside of the semiclassical regime as it violates the proper time condition \eqref{eq:time}.}
\label{Fig:Lambda-zero-k-zero}
\end{figure}
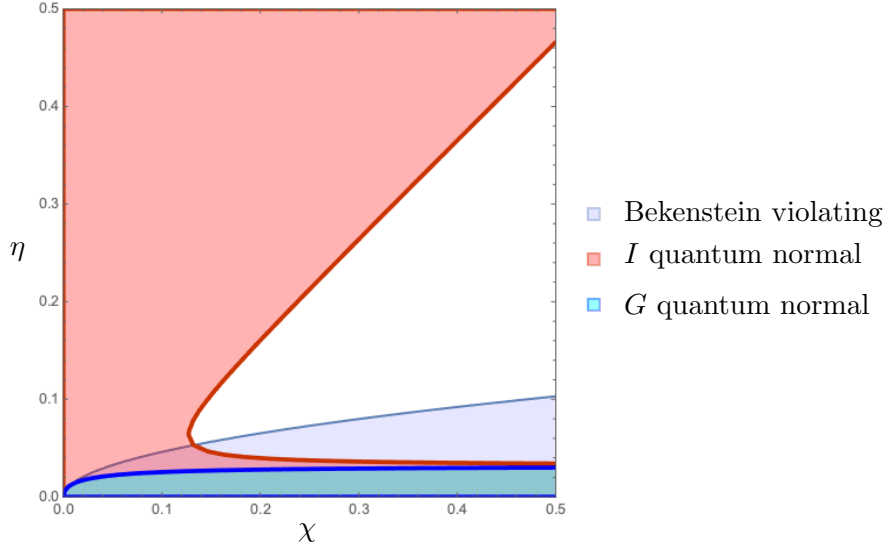

There is an overlapping region where all three island conditions are simultaneously satisfied and it is shown in teal in Figure \ref{Fig:Lambda-zero-k-zero}. However, this region is outside of the semiclassical regime of validity. We conclude that we do not have islands in this universe.

\subsection{Positive cosmological constant}
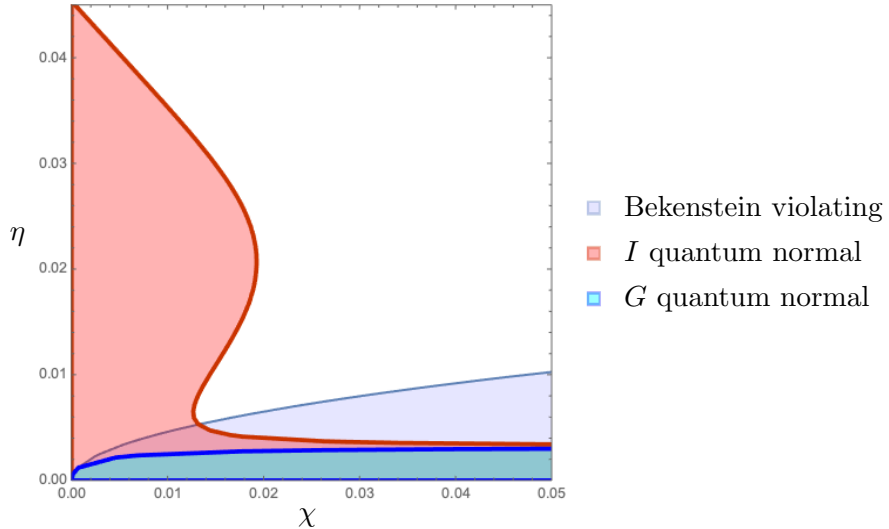
\begin{figure}[t]
 \centering
     \includegraphics[width=.5\linewidth]{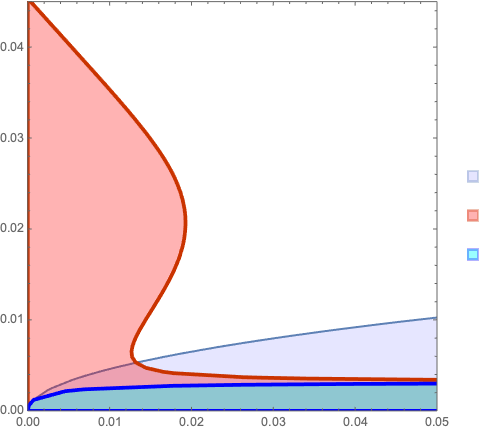}
      \begin{picture}(0,0)
\put(0,109){Bekenstein violating}
\put(0,92){$I$ quantum normal}
\put(0,75){$G$ quantum normal}
\put(-230,100){$\eta$}
\put(-122,-6){$\chi$}
\end{picture}
     \caption{Regions where the three island conditions are satisfied. We chose the numeric values $k=0$, $c_{th}=1$, $T_0=100$, and $\Lambda_0 G_N=0.01$. There is an overlapping region close to $\eta \approx 0$ where the proper time condition (\ref{eq:time}) is not satisfied.}
\label{Fig:Lambda-positive-k-zero}
\end{figure}

We proceed by turning on the cosmological constant and checking the three conditions. The Bekenstein bound is violated when 
\be
\chi_I \gtrsim \frac{3a(\eta)^2}{4G s_{\mathrm{th}}}~.
\ee
The region $I$ should be quantum normal. The outgoing condition  is always satisfied while the ingoing condition requires
\be 
\chi_I  \leq a(\eta)/a'(\eta).
\ee
The $G$ quantum normal condition implies 
\be
a(\eta)( \pm a'(\eta) \chi_I -a(\eta) )+ 2 G_N  s_{\mathrm{th}} \chi_I  \geq 0~.
\ee
Similarly as before, the overlap occurs outside of the semiclassical region. This part of the geometry is depicted in teal in Figure \ref{Fig:Lambda-positive-k-zero}.

\subsection{Negative cosmological constant}
Finally, we consider the case with negative cosmological constant. There is a recollapsing FRW universe as a  solution to the Friedmann equation (\ref{eq:Friedmann}). Importantly, a new ingredient of this cosmology is the existence of a time-symmetric slice. By solving  (\ref{eq:Friedmann}) at the time $\eta = \eta_0$ such that $a'(\eta_0)=0$, we get
\be\label{eq:a0flat}
a_0 =\left(\frac{8\pi G_N c_{\rm th }  T_0^4}{\Lambda_0} \right)^{\frac{1}{4}}~.
\ee
\subsubsection*{Island conditions}
At this particular time the Bekenstein bound is violated for values
\be\label{eq:BB_negative_k=0}
\chi_I \gtrsim
% \frac{3a_0^2}{4G_Ns_{th}}
\frac{9}{4 T_0}\left(\frac{\pi}{2 c_{\rm th} \Lambda_0 G_N} \right)^{1/2}~.
\ee
The ingoing and outgoing quantum normal conditions for region $I$ are reduced to one condition
\be
 \partial_{\chi} S_{\rm gen}(I) \geq 0~,
 \ee
which is always satisfied in this case. Similarly, the $G$ quantum normal condition becomes
\be
- \partial_{\chi} S_{\rm gen}(G) \geq 0 ~  \implies \chi_I \geq \frac{3}{2 T_0} \left(\frac{\pi}{2 c_{\rm th} \Lambda_0 G_N} \right)^{1/2}~,
\ee
 which is approximately the same result as the one we get from the first condition. Hence, there is triple overlap when \eqref{eq:BB_negative_k=0} is fulfilled.
 
\subsubsection*{Semiclassical regime}
We still have to check that these regions, along the time-symmetric slice, are in the semiclassical regime. The thermal length  (\ref{eq:beta}) and energy density  \eqref{eq:energy} conditions imply $\Lambda_0 G_N \ll 1$.
From (\ref{eq:sphere}), the size of the region has to be
\be
T_0 \chi_I \gg \left( \frac{\Lambda_0 G_N}{c_{\rm th}}\right)^{1/4}~,
\ee
which is a less restrictive condition than (\ref{eq:BB_negative_k=0}). The curvature radius condition (\ref{eq:curvature}) gives
\be
T_0 R_0 \gg \left(\frac{ \Lambda_0 G_N}{c_{\rm th}} \right)^{1/4}~.
\ee
All of the semiclassical conditions are satisfied and compatible with the island conditions in the overlapping region \eqref{eq:BB_negative_k=0}.
\subsection*{Conclusion}
\begin{figure}[t]
 \centering
     \includegraphics[width=.5\linewidth]{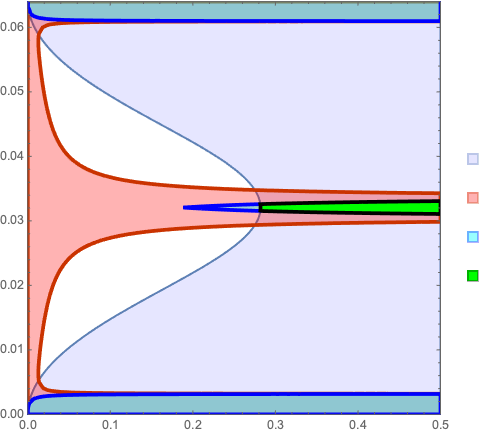}
      \begin{picture}(0,0)
\put(0,119){Bekenstein violating}
\put(0,101){$I$ quantum normal}
\put(0,84){$G$ quantum normal}
\put(0,66){Island}
\put(-230,100){$\eta$}
\put(-122,-6){$\chi$}
\end{picture}
\caption{Regions where the three island conditions are satisfied. We chose the numeric values, $k=0$, $c_{\rm th}=1$, $T_0=10$, and $\Lambda G_N=- 0.01$. We see that all conditions are simultaneously satisfied starting from the value $ \chi_I = 3a_0^2/4G_Ns_{\rm th}$. The island region is depicted in green. There are also overlapping regions at times where the solution $a(\eta)$ recollapses. However, they lie outside the semiclassical regime of validity given by the proper time condition (\ref{eq:time}).}
 \label{Fig:Lambda-negative-k-zero}
\end{figure}
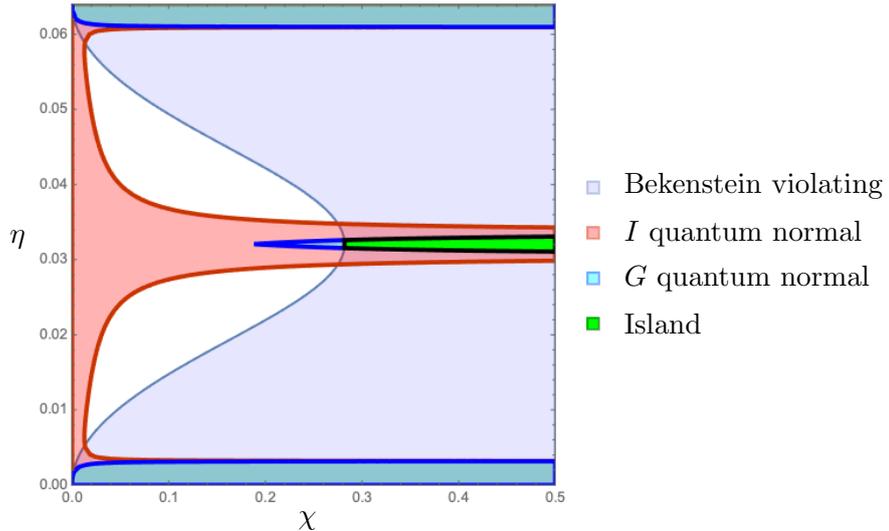

We conclude that there is an island region in the semiclassical regime for values that satisfy (\ref{eq:BB_negative_k=0}). This region is shown in green in Figure \ref{Fig:Lambda-negative-k-zero} together with the island conditions in the whole spacetime.
\section{Positive curvature}\label{positivecurvature}
\subsection{Zero cosmological constant}\label{Sec:lambdazero-closed}
We first consider the case where the vacuum energy density is vanishing and the FRW cosmology is supported by radiation and positive curvature. Without loss of generality, we can fix $k=1$. In these coordinates, $\chi \in [0,\pi R_0]$ is one of the angles that parametrize the $\mathcal{S}^3$. Solving (\ref{eq:Friedmann}), gives a scaling factor of the form 
\be\label{eq:afactor_Λ=0,k>0}
a(\eta) = \left( \frac{8 \pi G_N R_0^2 c_{\rm th} T_0^4}{3} \right)^{1/2}\sin\left( \frac{\eta}{R_0} \right)~.
\ee
\subsubsection*{Island conditions}
In order to look for island regions in this cosmology, we impose the three conditions to the region $I$ located at $\chi = \chi_I$ and $\eta = \eta_I$. 
The Bekenstein bound has the form
\be\label{eq:BB_flat_closed}
  2 \chi_I- R_0 \sin \left(\frac{2 \chi_I}{R_0} \right)  \gtrsim \frac{ a(\eta_I)^2}{s_{\rm th} G_N} \sin^2 \left( \frac{\chi_I}{R_0} \right) ~.\ee
In this equation we are basically comparing the comoving volume of the ${\mathcal{S}^3}$, with the area term $A(\partial I)/4G_N$. A natural place to look for island regions is the point where the volume is maximum and the area is very small. In fact, the ${\mathcal{S}}^3$ acquires its maximum size at $\chi_I = \pi R_0 $. Close to this particular value (\ref{eq:BB_flat_closed}) becomes
\be
\delta^2 \lesssim \frac{R_0}{T_0} \frac{1}{\sin \left(\frac{\eta_I}{R_0}\right)^2}~, ~~ \delta := \pi R_0 - \chi_I~.
\ee
Going to small times, $\eta_I/R_0 \ll 1$, so that the area term shrinks, we get
\be\label{eq:BBetasmall1}
\left(\frac{\delta}{R_0}\right)^2 \lesssim {R_0 \over T_0\eta_I^2}~.
\ee
Imposing that $I$ and $G$ should be quantum normal results in
\be\label{eq:Inormaletasmall}
T_0 \eta_I \leq \frac{1}{\pi}~.
\ee
This region, however, is non-semiclassical. We can see this by computing the proper time to the singularity and the thermal length. Both (\ref{eq:time}) and (\ref{eq:beta}) are valid when
\be
T_0 \eta_I \gg 1~,
\ee
which is clearly in conflict with (\ref{eq:Inormaletasmall}).

Therefore, there are no island regions close to the singularities located at $\eta_I \approx 0$ and $\eta_I \approx \pi$, and the place where the area term shrinks $\chi_I \approx \pi R_0$.
We now proceed to analyze the time-symmetric slice. At $\eta_I = \pi R_0 /2$, the scale factor is
\be
a_0 =\sqrt{\frac{8\pi c_{\rm th} G_N}{3}}R_0 T_0^2 ~.
\ee
Let us examine again the region close to  $\chi_I = \pi R_0$. Expanding (\ref{eq:BB_flat_closed}), we obtain
\be\label{eq:BB1_flat_closed}
\left(\frac{\delta}{R_0}\right)^2 \lesssim \frac{1}{R_0 T_0}~.
\ee
For $I$ to be quantum normal, we have
\be
\frac{\delta}{R_0} \geq \pi R_0 T_0~,
\ee
while the condition for region $G$ is always satisfied. 
We see that the three conditions are simultaneously true in the regime where $R_0 T_0 < 1$.
 \begin{figure}[t]
     \centering
     \includegraphics[width=.5\linewidth]{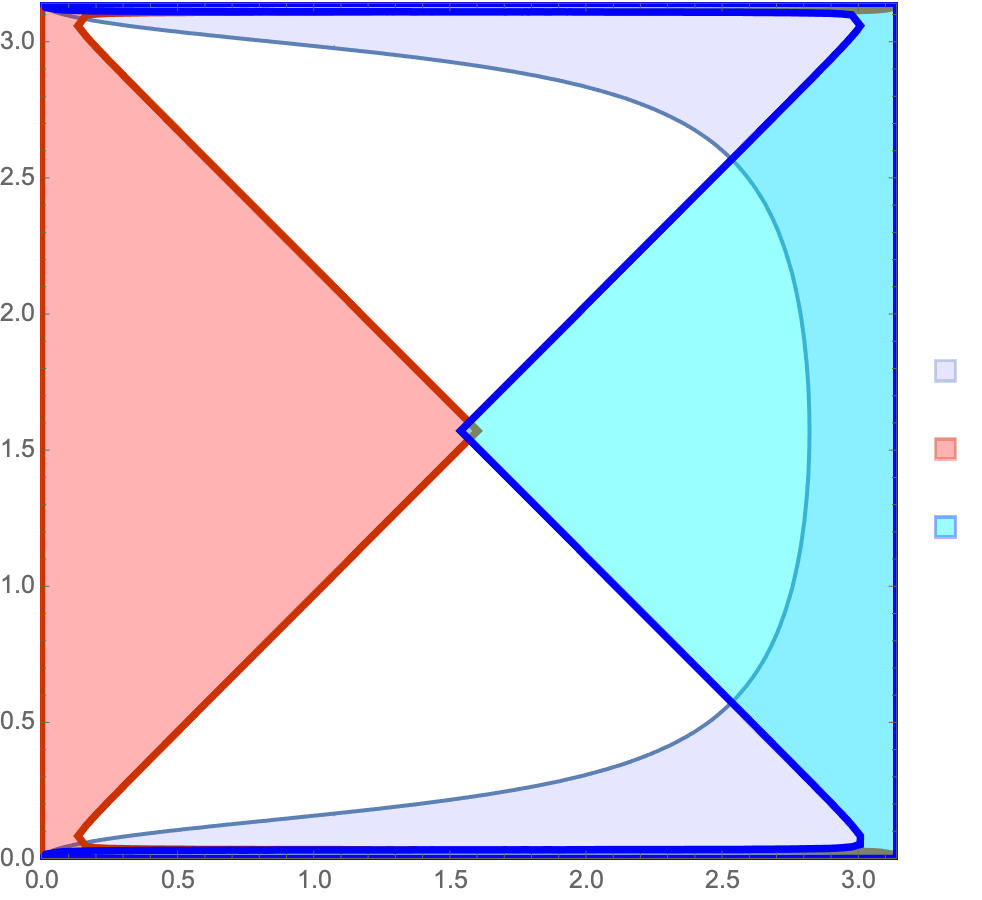}
     \begin{picture}(0,0)
\put(0,113){Bekenstein violating}
\put(0,97){$I$ quantum normal}
\put(0,79){$G$ quantum normal}
\put(-230,100){$\eta$}
\put(-122,-5){$\chi$}
\end{picture}
     \caption{Regions where the three conditions are satisfied. We chose the numeric values $k=1$, $\Lambda = 0$, $c_{th}=1$ and $R_0 T_0=10$ and $G_N=0.01$. There are overlapping regions close to the singularities, which are outside the semiclassical regime given by the proper time condition \eqref{eq:time}.}
\label{Fig:Λ=0,k>0}
\end{figure}
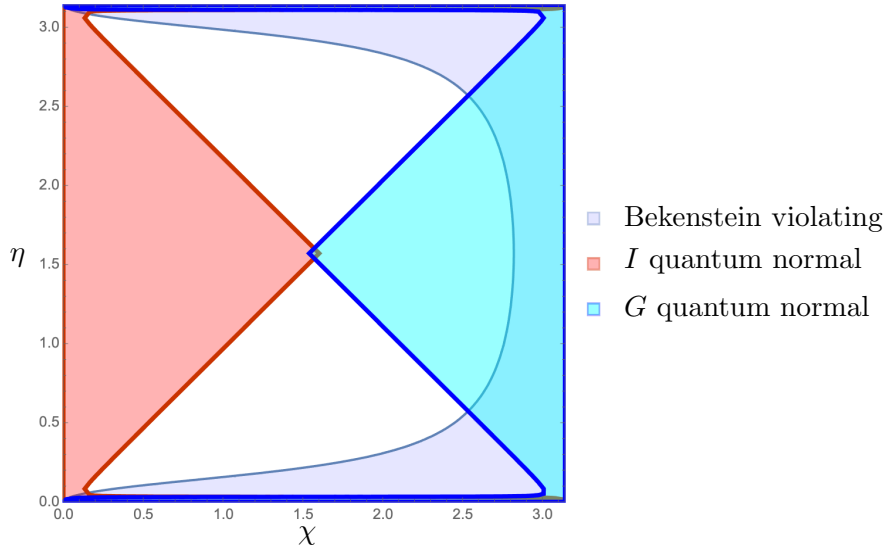
\subsubsection*{Semiclassical regime}
The semiclassical conditions for the thermal length (\ref{eq:beta}) and the energy density (\ref{eq:energy}) are satisfied for large values of the temperature
\be
R_0 T_0 \gg 1~. 
\ee
In this limit, the island conditions do not overlap and therefore there is no finite-size island at the time-symmetric slice. In Figure \ref{Fig:Λ=0,k>0}, we show the regions where the three conditions are valid in the semiclassical regime. As the temperature decreases into non-semiclassical values, the Bekenstein violating region progressively covers half of the spacetime, creating a triple overlap in the middle of the Penrose diagram. Moreover, there are overlapping regions close to the singularities outside the scope of the semiclassical analysis. 
\subsection*{Conclusion}
The only possible island is the the entire Cauchy slice.
\subsection{Positive cosmological constant}
As a warm-up example, we first neglect the density of radiation. This cosmology then corresponds to the thermal state in de Sitter space. Here, it is convenient to translate everything to the de Sitter radius $\ell_\mathrm{dS}$. In four dimensions, we have 
\be
\ell_\mathrm{dS} = \sqrt{\frac{3}{\Lambda_0}}~.
\ee
The Friedmann equation (\ref{eq:Friedmann}) reads
\be
\frac{1}{a(\eta)^2}\left(\frac{a'(\eta)}{a(\eta)} \right)^2 + \frac{1}{a(\eta)^2\ell_{\rm dS}^2} - \frac{\Lambda_0}{3}=0~,
\ee
and the solution is found to be
\be 
a(\eta) =\dfrac{1}{\cos({\frac{\eta}{\ell_{\rm dS}}})}~.
\ee
\subsection*{Island conditions}
\begin{figure}[t]
 \centering
     \includegraphics[width=.5\linewidth]{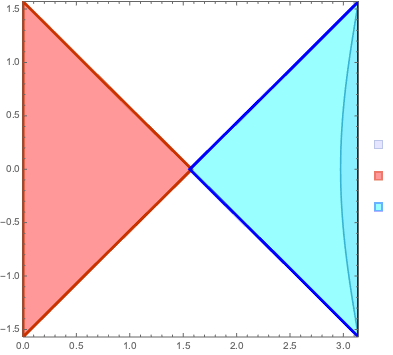}
      \begin{picture}(0,0)
\put(0,113){Bekenstein violating}
\put(0,95.5){$I$ quantum normal}
\put(0,78.5){$G$ quantum normal}
\put(-230,100){$\eta$}
\put(-122,-6){$\chi$}
\end{picture}
     \caption{Regions where the three conditions are satisfied. The numeric values are chosen to be $k=1$, $c_{th}=1$, $\Lambda_0 G_N=0.01$. There are no non-trivial island regions in thermal $dS_4$.}
     \label{fig: noradpositivem}
\end{figure}
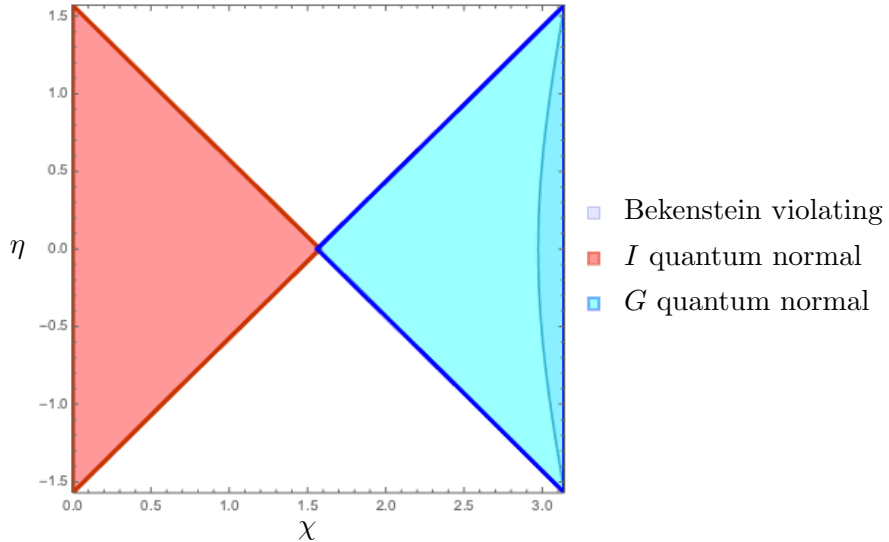

We now check the three necessary conditions. We restrict to the time-symmetric slice, where $a_0=1$. Here, the entropy density has the form $s_{\rm th} = 1/ \ell_{\rm dS}^3$. The Bekenstein bound then is
\be\label{eq:BBdS}
\frac{\pi}{G_N} \sin^2 \left( \frac{\chi_I}{\ell_{\rm dS}
} \right)\lesssim  \frac{1}{\ell_{\rm dS}^3}\left(2 \chi_I- \ell_{\rm dS}
 \sin \left(\frac{2 \chi_I}{\ell_{\rm dS}
} \right) \right)\,.
\ee
We notice that the first condition is satisfied at $\chi_I = \pi R_0$. If we expand close to that point, (\ref{eq:BBdS}) gives
\be 
\delta^2 \lesssim G_N ~, ~~ \delta :=  \pi R_0 - \chi_I~. 
\ee
This means that in order to satisfy the first island condition we have to go to the distance that is smaller than $l_P$ away from $\chi_I=\pi R_0$. Therefore, there are no islands smaller than the full Cauchy slice in the thermal $dS_4$. However, for completeness we continue the analysis. We consider the second condition which leads to
\be
\delta\geq \frac{\ell_{dS}^{3}}{G_N}~.
\ee
The third condition is always satisfied. In Figure \ref{fig: noradpositivem}, we show the three islands conditions in the semiclassical regime. We see that there are no overlapping regions.  
\subsection*{Conclusion}
The only possible island is the entire Cauchy slice.
\subsubsection{Adding radiation
}
We now add radiation. We focus on the time-symmetric slice
where (\ref{eq:Friedmann}) takes the form
\be
\label{eq521}
- \frac{8\pi G_N c_{\rm th}T_0^4}{3 a(\eta_0)^4} + \frac{1}{a(\eta_0)^2 R_0^2} - \frac{\Lambda_0}{3} =0~.
\ee
By solving \eqref{eq521} we obtain
\be 
\label{eqandrad}
a_0 = \sqrt{ 3  \pm \sqrt{9-32 \pi c_{\rm th}  \Lambda_0 G_N R_0^4 T_0^4} \over 2  R_0^2 \Lambda_0}~.
\ee
In order for this cosmology to have a time-symmetric solution, the following bound must be satisfied
\be
c_{\rm th} \Lambda_0 G_N (R_0 T_0)^4 \leq \frac{9}{32 \pi}~.
\ee
It is obvious that it is not possible to go to the high temperature limit since that would lead to a negative argument under the root in \eqref{eqandrad}. So, we focus on the low temperature limit.  
\subsubsection{Low temperatures}
In the low temperature limit, 
\be\label{eq:Tsmall}
R_0 T_0 \ll \frac{1}{(\Lambda_0 G_N)^{1/4}}~,
\ee 
there are two possible values for the scale factor, \ie, $a_0 \approx 1/R_0 \sqrt{\Lambda_0} $ and $a_0 \approx c_{\rm th} G_N^{1/2}R_0 T_0^2$. The latter is Planckian, so we do not explore this case further. The former represents the limit where the radiation becomes almost negligible and we check the island conditions below.

 \subsubsection*{Island conditions}
For scale factor equal to $a_0 \approx 1/R_0 \sqrt{\Lambda_0}$, we look for island regions other than the full Cauchy slice close to  $\chi = \pi R_0$.  As we explained before, in a closed universe, this is the region where the probability of violating the Bekenstein bound is the highest, since that is where the volume has its maximum value and the area its minimum.  The Bekenstein bound close to $\chi_I = \pi R_0$  leads to 
\be\label{eq:bb1}
\left(\frac{\delta}{R_0} \right)^2 \lesssim \frac{8 \pi}{9} c_{\rm th} (R_0 T_0)^3 \Lambda_0 G_N~.
\ee
Next, we consider the quantum normal condition for region $I$, which gives
\be\label{eq:inormal}
\frac{\delta}{R_0} \geq \frac{1}{c_{\rm th} (R_0 T_0)^3 \Lambda_0 G_N}~.
\ee
The third condition is always satisfied. Consequently, there is triple overlap when
\be\label{eq:ineq}
R_0 T_0 \gtrsim \frac{1}{\left(c_{\rm th} \Lambda_0 G_N \right)^{1/3}}~.
\ee
However, (\ref{eq:ineq}) together with (\ref{eq:Tsmall}) imply $\Lambda_0 G_N \gg 1$. The overlapping region is therefore outside the semiclassical regime. 
\subsubsection*{Semiclassical regime}
\begin{figure}[t]
     \centering
     \includegraphics[width=1.
     \linewidth]{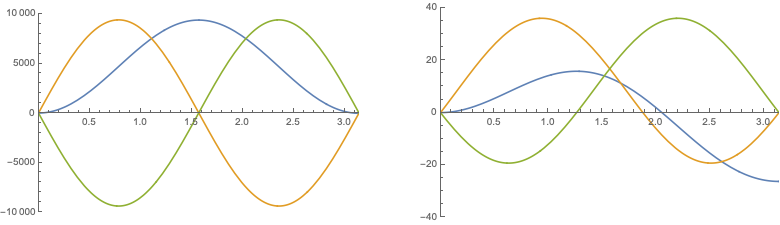}
     \begin{picture}(0,0)
\put(220,75){$\chi$}
\put(-10,75){$\chi$}
\end{picture}
     \caption{Three island conditions in the semiclassical regime for the two physical roots in (\ref{eqandrad}). The Bekenstein condition, $A(\partial I)/4 G_N - s_{\rm th}\widetilde{V}(I)$, is shown in blue, $I$ quantum normal condition in yellow and $G$ quantum normal condition in green. We chose the numeric values $k=1$, $c_{\rm th}=1$, $R_0 T_0=1$, and $\Lambda_0 G_N=0.001$. (Left) The Bekenstein bound is never satisfied, as $A(\partial I)/4 G_N - s_{\rm th}\widetilde{V}(I)$ is always positive. (Right) Bekenstein and $G$ conditions are both satisfied close to $\chi \approx \pi R$. In all cases, the $I$ and $G$ quantum normal conditions are mutually exclusive. The three islands conditions are never satisfied.}
\label{Fig:Roots}
  \end{figure}
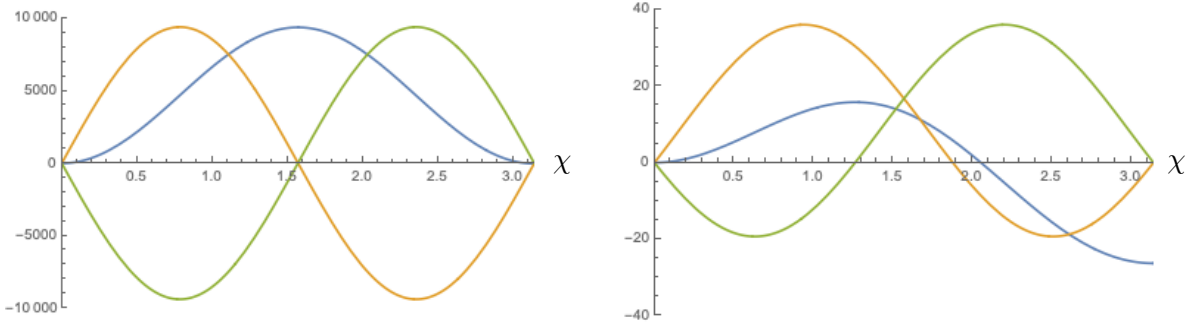

For semiclassical spacetimes, we know that $\Lambda_0 G_N \ll1$, as can be easily seen from (\ref{eq:curvature}). In Figure \ref{Fig:Roots}, we show three island conditions in the semiclassical limit. We see that for the two physical roots in (\ref{eqandrad}) the island conditions are never simultaneously satisfied. 
\subsection*{Conclusion} 
The only possible island is the entire Cauchy slice.

\subsection{Negative cosmological constant} 
We now turn to the case where $\Lambda = - \Lambda_0$, with $\Lambda_0 > 0$. At the time-symmetric slice (\ref{eq:Friedmann}) reduces to
\be
- \frac{8\pi G_N c_{\rm th} T_0^4}{3 a(\eta_0)^4} + \frac{1}{a(\eta_0)^2 R_0^2} + \frac{\Lambda_0}{3} =0~.
\ee
At this time, the scale factor acquires the value
\be\label{eq:afactor_Λ<0,k>0}
a_0 = \frac{1}{R_0 \sqrt{2 \Lambda_0}} \sqrt{-3 + \sqrt{9 +32 \pi c_{\rm th} G_NR_0^4 T_0^4 \Lambda_0}}~.
\ee
In the small temperature limit, $R_0 T_0 \ll 1/(\Lambda_0 G_N)^{1/4}$, (\ref{eq:afactor_Λ<0,k>0}) takes the form
\be
a_0 \approx \sqrt{\frac{8 \pi c_{th} G_N}{3}} R_0 T_0^2~. 
\ee
This is effectively the limit where the vacuum energy density is negligible and coincides with the analysis in Section \ref{Sec:lambdazero-closed} where there are no island regions. 
\subsubsection{High temperatures}
We now analyze the high temperature limit
\be\label{eq:highT}
R_0 T_0 \gg 1/(\Lambda_0 G_N)^{1/4}~. 
\ee
Here, (\ref{eq:afactor_Λ<0,k>0}) becomes 
\be\label{eq:aohighT}
a_0 \approx  \left( \frac{8\pi c_{\rm th} G_N T_0^4}{\Lambda_0}\right)^{1/4}~.
\ee
Importantly, in this limit the curvature contribution is negligible since the scale factor coincides with (\ref{eq:a0flat}). Naively, one might think that the best place to look for islands is where the volume is maximal as in the spatially flat case. However, for a closed universe, regions close to $\chi_I = \pi R_0/2$ are in fact anti-normal.
\subsubsection*{Island conditions}
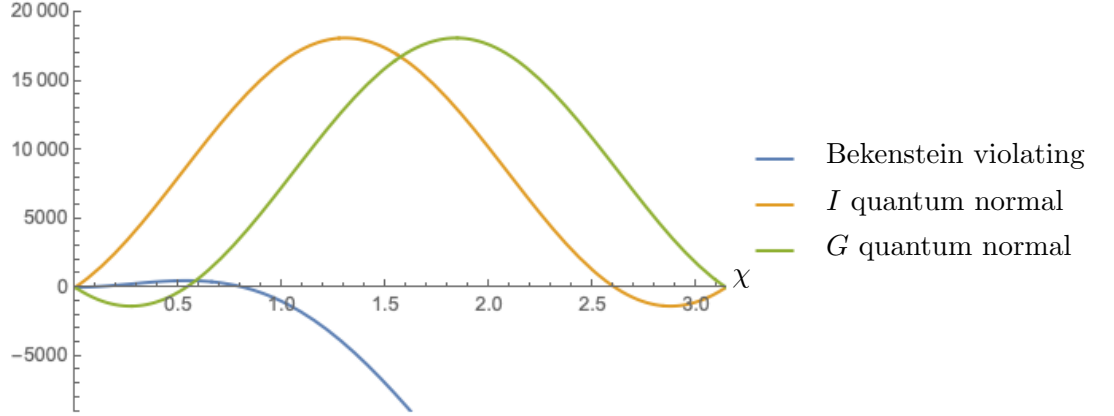
\begin{figure}[t]
     \centering
     \includegraphics[width=.7
     \linewidth]{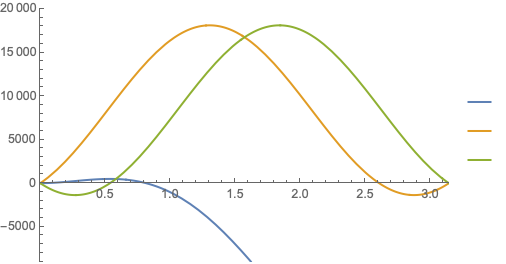}
     \begin{picture}(0,0)
\put(0,95){Bekenstein violating}
\put(0,77){$I$ quantum normal}
\put(0,60){$G$ quantum normal}
\put(-35,50){$\chi$}
%\put(-190,60){$\overbrace{~~~~~~~~~~~~~~~}$}
\end{picture}
     \caption{Three island conditions in the semiclassical regime along the time-symmetric slice. We use the scale factor (\ref{eq:aohighT}). The Bekenstein condition shown is $A(\partial I)/4 G_N - s_{th}\widetilde{V}(I)$, which is satisfied when this quantity is negative.  We chose the numeric values $k=1$, $c_{\rm th}=1$, $R_0 T_0=10$, and $\Lambda_0 G_N=0.1$. The three islands conditions are satisfied in the overlapping region $2 \delta$ around the value $\chi_I = \pi R_0/2$. }
\label{Fig:Lambdaneg}
  \end{figure}

It turns out that the best place to look for islands is at the middle of the $\mathcal{S}^3$ in the direction of $\chi$, \ie, $\chi_I =\pi R_0/2$. Let us first take values of $\chi_I$ such that $0 < \pi /2 - \chi_I/R_0 \ll 1$. For high temperatures, the Bekenstein bound reduces to the condition
\be\label{eq:BB_negative_closed}
\frac{\delta}{R_0} \lesssim  \frac{\pi}{4} - \frac{3}{8} \frac{1}{\sqrt{\pi c_{\rm th} \Lambda_0 G_N}} \frac{1}{R_0 T_0}~, ~~~ \delta := \frac{\pi R_0}{2} - \chi_I~.
\ee
The $G$ quantum normal condition to first order implies
\be\label{eq:G_negative_closed}
\frac{\delta}{R_0} \leq \frac{2}{3} \sqrt{\frac{2 c_{\rm th} \Lambda_0 G_N}{\pi}}R_0 T_0~,
\ee
while the condition for $I$ quantum normal is always satisfied. We see that both (\ref{eq:BB_negative_closed}) and (\ref{eq:G_negative_closed}) are satisfied when
\be\label{eq:overlapping}
R_0 T_0 > \frac{1}{\sqrt{c_{\rm th} \Lambda_0 G_N}}~.
\ee
This regime of parameters, where the three island conditions are satisfied, is consistent with the high temperature limit (\ref{eq:highT}).

\subsubsection*{Semiclassical regime}
We proceed to determine if the overlapping region is in the semiclassical regime. The condition for the proper time (\ref{eq:time}) is valid for times $T_0 \eta_I \gg (G_N \Lambda_0)^{1/4}$. The conditions (\ref{eq:beta}) and (\ref{eq:energy}) are satisfied when $\Lambda_0 G_N \ll 1$. The curvature condition (\ref{eq:curvature}) implies $R_0 T_0 \gg (G_N \Lambda_0)^{1/4}$. Finally, the sphere size condition (\ref{eq:sphere}), to leading order in the separation $\delta$, has the following form
\be
R_0 T_0 \gg \left(\frac{\Lambda_0 G_N}{8 \pi c_{\rm th}}\right)^{1/4} + \frac{1}{2}\left(\frac{\Lambda_0 G_N}{8 \pi  c_{\rm th}}\right)^{1/4} \left(\frac{\delta}{R_0}\right)^2 + \dots ~.
\ee
We see that all these conditions are compatible with (\ref{eq:overlapping}) in the high temperature limit (\ref{eq:highT}). We present the three island conditions in Figure \ref{Fig:Lambdaneg} for the time-symmetric slice.

\subsection*{Conclusion} There is a semi-classical region at the time-symmetric slice and around the half-sphere point, $\chi_I = \pi R_0 /2$, where the three island conditions are satisfied. Therefore, an island region appears when (\ref{eq:overlapping}) is satisfied together with $R_0 T_0 \gg 1$ and $\Lambda_0 G_N \ll 1$. A similar analysis follows for the values $\pi /2 - \chi_I/R_0 < 0$. In fact, there is a symmetric island region with respect to the half-sphere location. These regions are depicted in Figure \ref{Fig:Λ<0,k>0}. 
  
\section{Negative curvature}\label{negativecurvature}

\subsection{Zero or positive  cosmological constant}
These types of universes do not admit recollapsing solutions where $a'(\eta)\lvert_{\eta=\eta_0}=0$. Both of them expand forever.
 \begin{figure}[t]
     \centering
     \includegraphics[width=.5\linewidth]{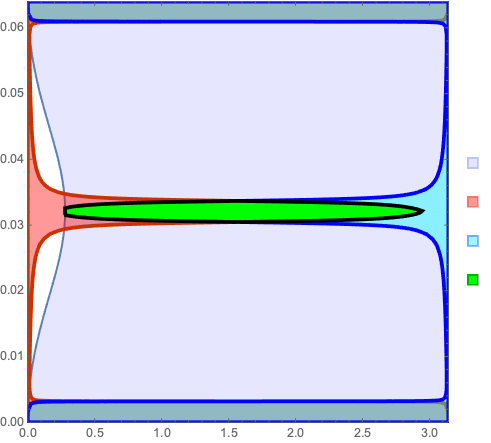}
     \begin{picture}(0,0)
\put(0,119.5){Bekenstein violating}
\put(0,102.5){$I$ quantum normal}
\put(0,86){$G$ quantum normal}
\put(0,69){Island}
\put(-230,100){$\eta$}
\put(-122,-6){$\chi$}
\end{picture}
     \caption{Regions where the three island conditions are satisfied in the semiclassical regime for a closed FRW cosmology. We chose the numeric values $k=1$, $c_{\rm th}=1$ and $R_0 T_0=100$ and $\Lambda_0 G_N=0.01$. There is an island region along the time-symmetric slice and around $\chi_I = \pi R_0/2$. There are also overlapping regions at times where the solution $a(\eta)$ recollapses. However, they lie outside the semiclassical regime.}
\label{Fig:Λ<0,k>0}
  \end{figure}
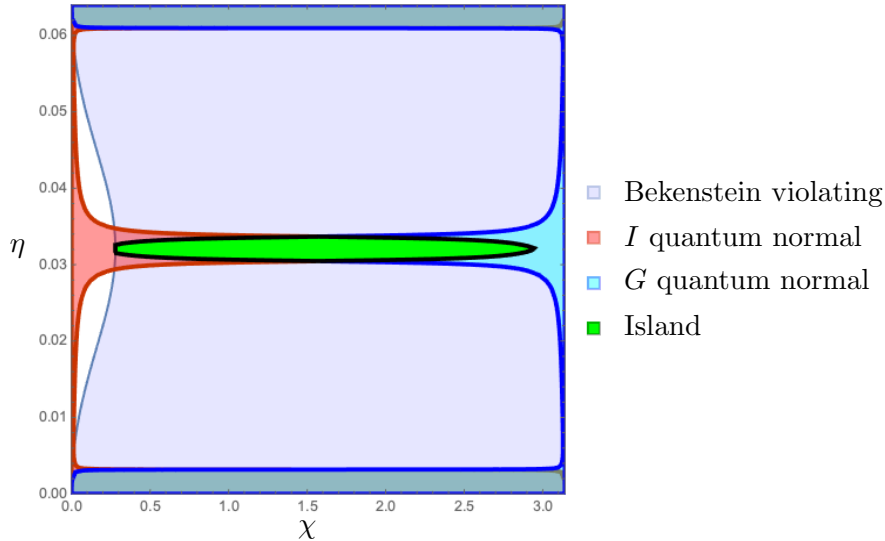

\subsection{Negative cosmological constant}
\begin{comment}
In this section we study the case of an radiation dominated open universe with negative cosmological constant. The form of the metric in this case is
\be
ds^2 = a^2(\eta) \left( -d\eta^2 + d \chi^2 + R^2 \sinh^2{\left(\frac{\chi}{R}\right)} d\Omega_2^2 \right) 
\ee
We take the values $k=-1$ and $\Lambda = - \Lambda_0$, and the Friedmann equation takes the form
\be
\label{ccrc}
\frac{1}{a(\eta)^2}\left(\frac{a'(\eta)}{a(\eta)} \right)^2 - \frac{8\pi G}{3}\epsilon - \frac{1}{a(\eta)^2 R^2} + \frac{\Lambda_0}{3} =0\,,
\ee
where $\epsilon=\epsilon_0/a(\eta)^4$. The solution to this equation is a complicated one, but we can solve it numerically.
\end{comment}
 We take the values $k=-1$ and $\Lambda = - \Lambda_0$. When we focus on the time-symmetric slice, \eqref{eq:Friedmann} simplifies to
\be
- \frac{8\pi G_N \epsilon_0}{3 a(\eta_0)^4 } - \frac{1}{a(\eta_0)^2 R_0^2} + \frac{\Lambda_0}{3} =0~.
\ee
Then the scale factor is 
\be
\label{lnegkneg}
a_0 = \frac{1}{R_0 \sqrt{2\Lambda_0}} \sqrt{3 + \sqrt{9 +32 \pi c_{\rm th} \Lambda_0 G_N R^4 T_0^4}}~.
\ee
\subsubsection{Low temperatures}
\begin{figure}[t]
     \centering
     \includegraphics[width=.7
     \linewidth]{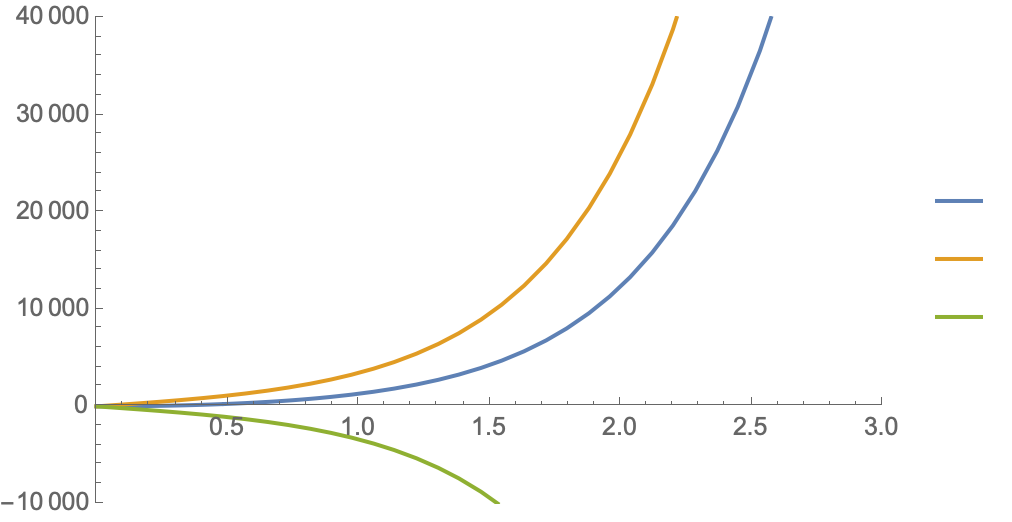}
     \begin{picture}(0,0)
\put(0,95){Bekenstein violating}
\put(0,77){$I$ quantum normal}
\put(0,60){$G$ quantum normal}
\put(-35,35){$\chi$}
\end{picture}
     \caption{Three island conditions in the semiclassical regime along the time-symmetric slice in the low temperature limit. We use the scale factor in (\ref{eq:a0smallT}). The Bekenstein condition shown is given by $A(\partial I)/4 G_N - s_{\rm th} \widetilde{V}(I)$, which is satisfied when this quantity is negative.  We chose the numeric values $k=-1$, $c_{th}=1$, $R_0 T_0=1$, and $\Lambda_0 G_N=0.01$. The $I$ quantum normal condition is always valid. The $G$ condition is never satisfed. The Bekenstein bound is never violated. }
\label{Fig:smallTopen}
  \end{figure}

First, in the low temperature limit $R_0 T_0\ll 1/\left( c_{\rm th} G_N\Lambda_0 \right)^{1/4}$, \eqref{lnegkneg} becomes 
\be\label{eq:a0smallT}
a_0 \approx \frac{1}{R_0}\sqrt{\frac{3}{\Lambda_0}}~.
\ee
We do not expect to find islands in this case, as this limit corresponds to having almost no radiation. In fact, when we analytically check the first condition, we end up in contradictions. Using \eqref{eq:a0smallT} for large values of $\chi/R_0 \gg 1$, we find that in order to violate the Bekenstein bound, we need 
\be
R_0 T_0\gtrsim \frac{1}{(c_{\rm th} G_N \Lambda_0)^{1/3}}~,
\ee
which is very large in the semiclassical regime. However, our initial assumption was that the temperature is very low. Similarly, for small values of $\chi_I/R \ll 1$, we find that the same condition (\ref{eq:BB}) requires 
\be
\frac{\chi_I}{R_0}\gtrsim \frac{1}{(R_0 T_0)^3 G \Lambda_0 }~,
\ee
which together with the small temperature limit imply that $\chi_I/R_0 \gg 1/(c_{\rm th} G_N \Lambda_0)^{1/4}$ outside of the regime of validity. Hence, the first condition is never satisfied and consequently it is not possible to have islands. In Figure \ref{Fig:smallTopen}, we show the three conditions in the small temperature limit. 

\subsubsection{High temperatures}
We now look at the high temperature limit $ R_0 T_0 \gg 1/\left(c_{\rm th} G_N\Lambda_0 \right)^{1/4}$. Here, \eqref{lnegkneg} takes the form
\be\label{eq:a0largeT}
a_0 =  \left( \frac{8 \pi G_N  c_{th} T_0^4}{\Lambda_0}\right)^{1/4}\,,
\ee
which corresponds to having negligible curvature at the turnaround time. 

 \subsection*{Island conditions}
For small $\chi_I/R_0 \ll 1$, we find for the Bekenstein bound 
\be
\label{xIlnegkneg}
\frac{\chi_I}{R_0} \gtrsim \frac{9}{4 R_0 T_0}\left(\frac{\pi}{2 c_{\rm th} \Lambda_0 G_N} \right)^{1/2}\,.
\ee
For large $\chi_I/R_0 \gg 1$, we get
\be
\label{T0lnegkneg}
R_0 T_0\gtrsim \frac{3}{2}\left( \frac{\pi}{2c_{\rm th}G_N\Lambda_0}\right)^{1/2}\,.
\ee
The second condition that requires $I$ to be quantum normal is always satisfied at the time-symmetric slice, and the third results to the same inequalities up to small order one numbers as the first condition. Therefore, there is an overlapping region when (\ref{xIlnegkneg}) is valid.

\subsection*{Semiclassical regime}
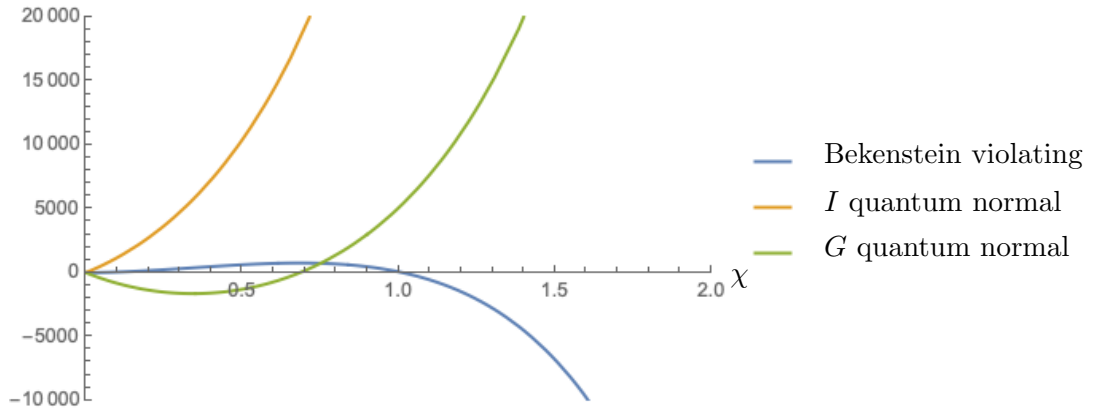
\begin{figure}[t]
     \centering
     \includegraphics[width=.7
     \linewidth]{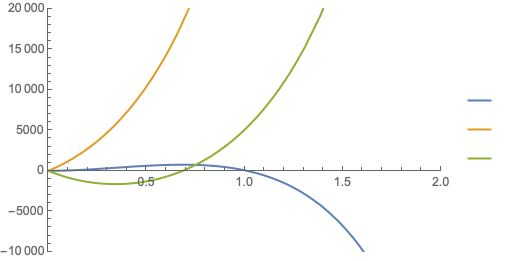}
     \begin{picture}(0,0)
\put(0,95){Bekenstein violating}
\put(0,77){$I$ quantum normal}
\put(0,60){$G$ quantum normal}
\put(-35,50){$\chi$}
\end{picture}
     \caption{Three island conditions in the semiclassical regime along the time-symmetric slice in the high temperature limit. We use the scale factor in (\ref{eq:a0largeT}). The Bekenstein condition shown is given by $A(\partial I)/4 G_N - s_{\rm th}\widetilde{V}(I)$, which is satisfied when this quantity is negative.  We chose the numeric values $k=-1$, $c_{\rm th}=1$, $R_0 T_0=10$, and $\Lambda_0 G_N=0.1$. The $I$ quantum normal condition is always valid. The three island conditions are satisfied for values in \eqref{xIlnegkneg}.}
     \label{Fig:open}
  \end{figure}

We also check that the semiclassical conditions at the time-symmetric slice are satisfied. The conditions
for the thermal length \eqref{eq:beta} and  energy density \eqref{eq:energy} are obeyed for $G_N \Lambda_0 \ll 1$, which are true for reasonable spacetimes. Moreover, we require that the size of the $\cS^2$ sphere is bigger than the Planck length. For large $\chi/R_0 \gg1$ the condition is automatically satisfied. For small $\chi/R_0 \ll 1$, we have 
\be
\frac{\chi}{R_0} \gg \frac{1}{R_0 T_0}\left(\frac{G_N \Lambda_0}{c_{\rm th}}\right)^{1/4}\,,
\ee
which is compatible with \eqref{xIlnegkneg}. Finally, we check the radius of curvature condition (\ref{eq:curvature}) 
\be
R_0 T_0 \gg \left(\frac{G_N \Lambda_0}{c_{\rm th}}\right)^{1/4}\,,
\ee
which is again compatible with \eqref{T0lnegkneg}. Therefore, when we combine all the conditions, we get an elongated teardrop region, (\ref{xIlnegkneg}). In Figure \ref{Fig:open}, we show three island conditions in the semiclassical limit. 

\subsection*{Conclusion}
There is a semiclassical region at the time-symmetric slice where the three island conditions are satisfied. Therefore, the existence of islands is possible for the values in \eqref{T0lnegkneg} together with $R_0 T_0 \gg 1$, and $\Lambda_0 G_N \ll 1$. This region is shown in green in Figure \ref{Fig:Λ<0,k<0}.
 
\section{Discussion}\label{Sec:Discussion}
 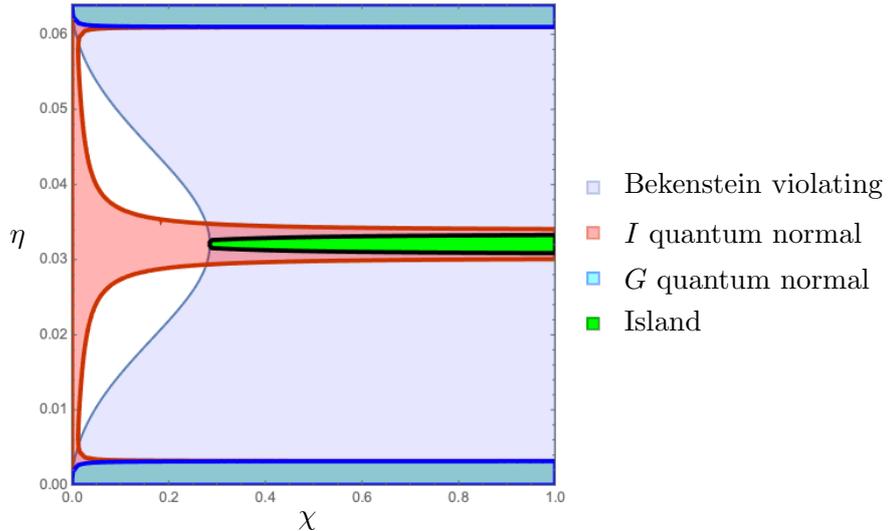
\begin{figure}[t]
 \centering
     \includegraphics[width=.5\linewidth]{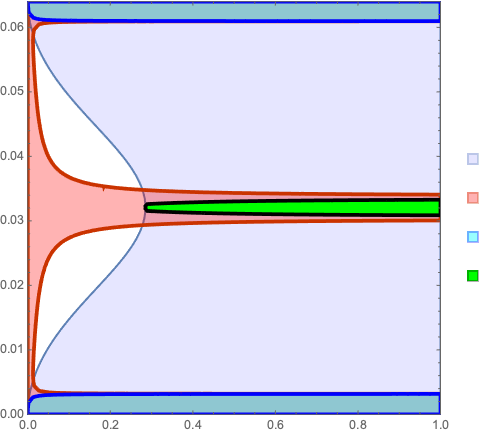}
      \begin{picture}(0,0)
\put(0,119){Bekenstein violating}
\put(0,100){$I$ quantum normal}
\put(0,83){$G$ quantum normal}
\put(0,67){Island}
\put(-230,100){$\eta$}
\put(-122,-6){$\chi$}
\end{picture}
     \caption{Regions where the three island conditions are satisfied in the semiclassical
regime for an open FRW cosmology. We chose the numeric values $k=-1$, $c_{\rm th}=1$, $R_0 T_0=100$, and $\Lambda_0 G_N=0.01$. There is an island region along the time-symmetric slice that extends to infinity. There are also overlapping regions at times where the solution $a(\eta)$ recollapses. However, they lie outside the semiclassical regime.}
\label{Fig:Λ<0,k<0}
\end{figure}
In this paper, we studied the possible existence of islands in FRW cosmologies supported by radiation, curvature, and cosmological constant. To this end, we applied the three necessary conditions to subregions of the spacetime $I$ and $G$, together with the semiclassical conditions introduced in Section \ref{framework}. We found that the key element for the existence of non-trivial islands is a negative cosmological constant. In the case of a closed universe, there is an island region around the half-sphere point located at $\chi_I = \pi R_0/2$. This island is finite in size and is qualitatively different from the island found in \cite{Hartman:2020khs}. For open universes, an island region shows up for large enough radius and extends all the way to infinity. By studying the spacetime at the time-symmetric slice, we provided analytic evidence for the existence of these islands in the high temperature limit where the spatial curvature is negligible. We also performed a numerical analysis in the entire spacetime.

It turns out that having an FRW cosmology with a time-symmetric slice is not sufficient for the existence of islands smaller than the whole Cauchy slice. For example, closed de Sitter and a recollapsing universe with $\Lambda=0$ and closed slicing have time-symmetric slices, but the only possible island is the full Cauchy slice.

The analysis carried out in this paper is valid for more general class of cosmologies. We expect to find islands in cosmologies with $\Lambda<0$ for which the effective potential $V_{\rm eff}(a)$ vanishes at the turning point $a'(\eta)=0$. For example, we expect that adding ordinary matter would not change the conclusions of this paper.

 There are a few exciting avenues of research related to our work that are worth exploring. Since currently our universe is undergoing another period of inflation, it would be interesting to study if islands are relevant in the context of inflation. This can be modeled
 by bubbles of false vacuum where an inflating region forms behind the horizon \cite{Coleman:1980aw, Farhi:1986ty,Freivogel:2005qh}. It is worth understanding whether inflating regions are encoded in non-gravitating systems and if their formation is allowed in our universe. Additionally, it would also be interesting to understand the implications of our work in 
  the case of eternally inflating multiverse studied in \cite{Langhoff:2021uct} where islands have been shown to appear in collapsing AdS bubbles. Furthermore, islands in Jackiw–Teitelboim 
 de Sitter multiverse
 have been studied in \cite{Aguilar-Gutierrez:2021bns} where they appeared in the crunching regions. Another direction of inquiry would be to explore the existence of islands in more general cosmologies in the spirit of \cite{Aguilar-Gutierrez:2021bns}.
\acknowledgments
It is a pleasure to thank Ben Freivogel and Andrew Rolph for initial collaboration, fruitful discussions, and useful comments. We also thank Tarek Anous, Carlos Duaso Pueyo, Facundo Rost, Jeremy van der Heijden, Amanda van Hemert, and Erik Verlinde for helpful discussions. RE and DN are supported by the ERC Consolidator Grant QUANTIVIOL. BN is supported by the Spinoza Grant of the Dutch Science Organisation (NWO).
This work is supported by the Delta ITP consortium, a program of the Netherlands Organisation for Scientific Research (NWO) that is funded
by the Dutch Ministry of Education, Culture and Science (OCW).

\bibliographystyle{JHEP}
\bibliography{islands.bib}

\providecommand{\href}[2]{#2}\begingroup\raggedright\begin{thebibliography}{10}

\bibitem{Hartman:2020khs}
T.~Hartman, Y.~Jiang and E.~Shaghoulian, \emph{{Islands in cosmology}},
  \href{https://doi.org/10.1007/JHEP11(2020)111}{\emph{JHEP} {\bfseries 11}
  (2020) 111} [\href{https://arxiv.org/abs/2008.01022}{{\ttfamily
  2008.01022}}].

\bibitem{Hawking:1975vcx}
S.W.~Hawking, \emph{{Particle Creation by Black Holes}},
  \href{https://doi.org/10.1007/BF02345020}{\emph{Commun. Math. Phys.}
  {\bfseries 43} (1975) 199}.

\bibitem{Hawking:1976ra}
S.W.~Hawking, \emph{{Breakdown of Predictability in Gravitational Collapse}},
  \href{https://doi.org/10.1103/PhysRevD.14.2460}{\emph{Phys. Rev. D}
  {\bfseries 14} (1976) 2460}.

\bibitem{Page:1993wv}
D.N.~Page, \emph{{Information in black hole radiation}},
  \href{https://doi.org/10.1103/PhysRevLett.71.3743}{\emph{Phys. Rev. Lett.}
  {\bfseries 71} (1993) 3743}
  [\href{https://arxiv.org/abs/hep-th/9306083}{{\ttfamily hep-th/9306083}}].

\bibitem{Almheiri:2019psf}
A.~Almheiri, N.~Engelhardt, D.~Marolf and H.~Maxfield, \emph{{The entropy of
  bulk quantum fields and the entanglement wedge of an evaporating black
  hole}}, \href{https://doi.org/10.1007/JHEP12(2019)063}{\emph{JHEP} {\bfseries
  12} (2019) 063} [\href{https://arxiv.org/abs/1905.08762}{{\ttfamily
  1905.08762}}].

\bibitem{Penington:2019npb}
G.~Penington, \emph{{Entanglement Wedge Reconstruction and the Information
  Paradox}}, \href{https://doi.org/10.1007/JHEP09(2020)002}{\emph{JHEP}
  {\bfseries 09} (2020) 002}
  [\href{https://arxiv.org/abs/1905.08255}{{\ttfamily 1905.08255}}].

\bibitem{Almheiri:2019hni}
A.~Almheiri, R.~Mahajan, J.~Maldacena and Y.~Zhao, \emph{{The Page curve of
  Hawking radiation from semiclassical geometry}},
  \href{https://doi.org/10.1007/JHEP03(2020)149}{\emph{JHEP} {\bfseries 03}
  (2020) 149} [\href{https://arxiv.org/abs/1908.10996}{{\ttfamily
  1908.10996}}].

\bibitem{Almheiri:2019qdq}
A.~Almheiri, T.~Hartman, J.~Maldacena, E.~Shaghoulian and A.~Tajdini,
  \emph{{Replica Wormholes and the Entropy of Hawking Radiation}},
  \href{https://doi.org/10.1007/JHEP05(2020)013}{\emph{JHEP} {\bfseries 05}
  (2020) 013} [\href{https://arxiv.org/abs/1911.12333}{{\ttfamily
  1911.12333}}].

\bibitem{Penington:2019kki}
G.~Penington, S.H.~Shenker, D.~Stanford and Z.~Yang, \emph{{Replica wormholes
  and the black hole interior}},
  \href{https://arxiv.org/abs/1911.11977}{{\ttfamily 1911.11977}}.

\bibitem{Almheiri:2020cfm}
A.~Almheiri, T.~Hartman, J.~Maldacena, E.~Shaghoulian and A.~Tajdini,
  \emph{{The entropy of Hawking radiation}},
  \href{https://doi.org/10.1103/RevModPhys.93.035002}{\emph{Rev. Mod. Phys.}
  {\bfseries 93} (2021) 035002}
  [\href{https://arxiv.org/abs/2006.06872}{{\ttfamily 2006.06872}}].

\bibitem{Lewkowycz:2013nqa}
A.~Lewkowycz and J.~Maldacena, \emph{{Generalized gravitational entropy}},
  \href{https://doi.org/10.1007/JHEP08(2013)090}{\emph{JHEP} {\bfseries 08}
  (2013) 090} [\href{https://arxiv.org/abs/1304.4926}{{\ttfamily 1304.4926}}].

\bibitem{Ryu:2006bv}
S.~Ryu and T.~Takayanagi, \emph{{Holographic derivation of entanglement entropy
  from AdS/CFT}},
  \href{https://doi.org/10.1103/PhysRevLett.96.181602}{\emph{Phys. Rev. Lett.}
  {\bfseries 96} (2006) 181602}
  [\href{https://arxiv.org/abs/hep-th/0603001}{{\ttfamily hep-th/0603001}}].

\bibitem{Hubeny:2007xt}
V.E.~Hubeny, M.~Rangamani and T.~Takayanagi, \emph{{A Covariant holographic
  entanglement entropy proposal}},
  \href{https://doi.org/10.1088/1126-6708/2007/07/062}{\emph{JHEP} {\bfseries
  07} (2007) 062} [\href{https://arxiv.org/abs/0705.0016}{{\ttfamily
  0705.0016}}].

\bibitem{Faulkner:2013ana}
T.~Faulkner, A.~Lewkowycz and J.~Maldacena, \emph{{Quantum corrections to
  holographic entanglement entropy}},
  \href{https://doi.org/10.1007/JHEP11(2013)074}{\emph{JHEP} {\bfseries 11}
  (2013) 074} [\href{https://arxiv.org/abs/1307.2892}{{\ttfamily 1307.2892}}].

\bibitem{Engelhardt:2014gca}
N.~Engelhardt and A.C.~Wall, \emph{{Quantum Extremal Surfaces: Holographic
  Entanglement Entropy beyond the Classical Regime}},
  \href{https://doi.org/10.1007/JHEP01(2015)073}{\emph{JHEP} {\bfseries 01}
  (2015) 073} [\href{https://arxiv.org/abs/1408.3203}{{\ttfamily 1408.3203}}].

\bibitem{Dong:2016hjy}
X.~Dong, A.~Lewkowycz and M.~Rangamani, \emph{{Deriving covariant holographic
  entanglement}}, \href{https://doi.org/10.1007/JHEP11(2016)028}{\emph{JHEP}
  {\bfseries 11} (2016) 028}
  [\href{https://arxiv.org/abs/1607.07506}{{\ttfamily 1607.07506}}].

\bibitem{Dong:2016fnf}
X.~Dong, \emph{{The Gravity Dual of Renyi Entropy}},
  \href{https://doi.org/10.1038/ncomms12472}{\emph{Nature Commun.} {\bfseries
  7} (2016) 12472} [\href{https://arxiv.org/abs/1601.06788}{{\ttfamily
  1601.06788}}].

\bibitem{Almheiri:2019yqk}
A.~Almheiri, R.~Mahajan and J.~Maldacena, \emph{{Islands outside the horizon}},
   \href{https://arxiv.org/abs/1910.11077}{{\ttfamily 1910.11077}}.

\bibitem{Almheiri:2019psy}
A.~Almheiri, R.~Mahajan and J.E.~Santos, \emph{{Entanglement islands in higher
  dimensions}},
  \href{https://doi.org/10.21468/SciPostPhys.9.1.001}{\emph{SciPost Phys.}
  {\bfseries 9} (2020) 001} [\href{https://arxiv.org/abs/1911.09666}{{\ttfamily
  1911.09666}}].

\bibitem{Chen:2019uhq}
H.Z.~Chen, Z.~Fisher, J.~Hernandez, R.C.~Myers and S.-M.~Ruan,
  \emph{{Information Flow in Black Hole Evaporation}},
  \href{https://doi.org/10.1007/JHEP03(2020)152}{\emph{JHEP} {\bfseries 03}
  (2020) 152} [\href{https://arxiv.org/abs/1911.03402}{{\ttfamily
  1911.03402}}].

\bibitem{Rozali:2019day}
M.~Rozali, J.~Sully, M.~Van~Raamsdonk, C.~Waddell and D.~Wakeham,
  \emph{{Information radiation in BCFT models of black holes}},
  \href{https://doi.org/10.1007/JHEP05(2020)004}{\emph{JHEP} {\bfseries 05}
  (2020) 004} [\href{https://arxiv.org/abs/1910.12836}{{\ttfamily
  1910.12836}}].

\bibitem{Chen:2020uac}
H.Z.~Chen, R.C.~Myers, D.~Neuenfeld, I.A.~Reyes and J.~Sandor, \emph{{Quantum
  Extremal Islands Made Easy, Part I: Entanglement on the Brane}},
  \href{https://doi.org/10.1007/JHEP10(2020)166}{\emph{JHEP} {\bfseries 10}
  (2020) 166} [\href{https://arxiv.org/abs/2006.04851}{{\ttfamily
  2006.04851}}].

\bibitem{Balasubramanian:2020coy}
V.~Balasubramanian, A.~Kar and T.~Ugajin, \emph{{Entanglement between two
  disjoint universes}},
  \href{https://doi.org/10.1007/JHEP02(2021)136}{\emph{JHEP} {\bfseries 02}
  (2021) 136} [\href{https://arxiv.org/abs/2008.05274}{{\ttfamily
  2008.05274}}].

\bibitem{Balasubramanian:2020xqf}
V.~Balasubramanian, A.~Kar and T.~Ugajin, \emph{{Islands in de Sitter space}},
  \href{https://doi.org/10.1007/JHEP02(2021)072}{\emph{JHEP} {\bfseries 02}
  (2021) 072} [\href{https://arxiv.org/abs/2008.05275}{{\ttfamily
  2008.05275}}].

\bibitem{Bousso:2020kmy}
R.~Bousso and E.~Wildenhain, \emph{{Gravity/ensemble duality}},
  \href{https://doi.org/10.1103/PhysRevD.102.066005}{\emph{Phys. Rev. D}
  {\bfseries 102} (2020) 066005}
  [\href{https://arxiv.org/abs/2006.16289}{{\ttfamily 2006.16289}}].

\bibitem{Goto:2020wnk}
K.~Goto, T.~Hartman and A.~Tajdini, \emph{{Replica wormholes for an evaporating
  2D black hole}}, \href{https://doi.org/10.1007/JHEP04(2021)289}{\emph{JHEP}
  {\bfseries 04} (2021) 289}
  [\href{https://arxiv.org/abs/2011.09043}{{\ttfamily 2011.09043}}].

\bibitem{Bousso:2021sji}
R.~Bousso and A.~Shahbazi-Moghaddam, \emph{{Island Finder and Entropy Bound}},
  \href{https://doi.org/10.1103/PhysRevD.103.106005}{\emph{Phys. Rev. D}
  {\bfseries 103} (2021) 106005}
  [\href{https://arxiv.org/abs/2101.11648}{{\ttfamily 2101.11648}}].

\bibitem{Chen:2020tes}
Y.~Chen, V.~Gorbenko and J.~Maldacena, \emph{{Bra-ket wormholes in
  gravitationally prepared states}},
  \href{https://doi.org/10.1007/JHEP02(2021)009}{\emph{JHEP} {\bfseries 02}
  (2021) 009} [\href{https://arxiv.org/abs/2007.16091}{{\ttfamily
  2007.16091}}].

\bibitem{Maldacena:1997re}
J.M.~Maldacena, \emph{{The Large N limit of superconformal field theories and
  supergravity}}, \href{https://doi.org/10.1023/A:1026654312961}{\emph{Adv.
  Theor. Math. Phys.} {\bfseries 2} (1998) 231}
  [\href{https://arxiv.org/abs/hep-th/9711200}{{\ttfamily hep-th/9711200}}].

\bibitem{Bousso:2022gth}
R.~Bousso and E.~Wildenhain, \emph{{Islands in Closed and Open Universes}},
  \href{https://arxiv.org/abs/2202.05278}{{\ttfamily 2202.05278}}.

\bibitem{Coleman:1980aw}
S.R.~Coleman and F.~De~Luccia, \emph{{Gravitational Effects on and of Vacuum
  Decay}}, \href{https://doi.org/10.1103/PhysRevD.21.3305}{\emph{Phys. Rev. D}
  {\bfseries 21} (1980) 3305}.

\bibitem{Farhi:1986ty}
E.~Farhi and A.H.~Guth, \emph{{An Obstacle to Creating a Universe in the
  Laboratory}}, \href{https://doi.org/10.1016/0370-2693(87)90429-1}{\emph{Phys.
  Lett. B} {\bfseries 183} (1987) 149}.

\bibitem{Freivogel:2005qh}
B.~Freivogel, V.E.~Hubeny, A.~Maloney, R.C.~Myers, M.~Rangamani and S.~Shenker,
  \emph{{Inflation in AdS/CFT}},
  \href{https://doi.org/10.1088/1126-6708/2006/03/007}{\emph{JHEP} {\bfseries
  03} (2006) 007} [\href{https://arxiv.org/abs/hep-th/0510046}{{\ttfamily
  hep-th/0510046}}].

\bibitem{Langhoff:2021uct}
K.~Langhoff, C.~Murdia and Y.~Nomura, \emph{{Multiverse in an inverted
  island}}, \href{https://doi.org/10.1103/PhysRevD.104.086007}{\emph{Phys. Rev.
  D} {\bfseries 104} (2021) 086007}
  [\href{https://arxiv.org/abs/2106.05271}{{\ttfamily 2106.05271}}].

\bibitem{Aguilar-Gutierrez:2021bns}
S.E.~Aguilar-Gutierrez, A.~Chatwin-Davies, T.~Hertog, N.~Pinzani-Fokeeva and
  B.~Robinson, \emph{{Islands in Multiverse Models}},
  \href{https://arxiv.org/abs/2108.01278}{{\ttfamily 2108.01278}}.

\end{thebibliography}\endgroup

\end{document}